\documentclass[twoside,11pt]{article}
\usepackage{jair, theapa, rawfonts,url, graphicx}
\usepackage{array,multirow}
\usepackage{makecell}
\usepackage{tabularx}
\usepackage{booktabs}

\newcolumntype{P}[1]{>{\centering\arraybackslash}p{#1}}

\jairheading{1}{2016}{1-15}{6/91}{9/91}
\ShortHeadings{AI and Labour}
{S. Samothrakis}
\firstpageno{1}

\begin{document}

\title{Viewpoint: Artificial Intelligence and Labour}

\author{{\name Spyridon Samothrakis \email ssamot@essex.ac.uk \\		
    \addr Institute of Analytics and Data Science \\
       Wivenhoe Park\\
       Colchester CO4 3SQ\\
       Essex\\
       U.K. \\
     }
}

\newcommand{\rot}[2]{\rotatebox[origin=c]{90}{\parbox[c]{#1}{\centering #2}}}
\newcolumntype{C}[1]{>{\centering\let\newline\\\arraybackslash\hspace{0pt}}m{#1}}


\maketitle

\begin{abstract}
The welfare of modern societies has been intrinsically linked to wage labour. With some exceptions, the modern human has to sell her labour-power to be able reproduce biologically and socially. Thus, a lingering fear of technological unemployment features predominately as a theme among Artificial Intelligence researchers. In this short paper we show that, if past trends are anything to go by, this fear is irrational. On the contrary, we argue that the main problem humanity will be facing is the normalisation of extremely long working hours. 
\end{abstract}

\section{Introduction}
The most common view on the impact of intelligent machines\footnote{We will use the ``intelligent machines'' to include all kinds of intelligent algorithms as well as their physical manifestations.} is best captured by ex-Sun Microsystems's  ex-chief scientist paper~\cite{joy2000future}. Joy quotes Kaczynski’s ``Unabomber Manifesto''~\cite{kaczynski2005unabomber}, which works from the premise that the rise of intelligent machines will result either in a vast portion of humanity being exterminated or being relegated to ``pets''\footnote{The pet theory is captured nicely in Ian Banks' ``Culture'' series.} and tries to identify an alternative future for mankind. Kaczynski's positions of the lower strata of humanity being exterminated is dubbed ``exterminism '' by Frase~\cite{FourF13:online} and understood as a possible outcome  (among a total of four) of the future of humanity. The argument on why something like that would happen more or less goes along these lines; since machines can do a man's work, a portion of humanity will be superfluous. That portion of humanity will gradually and naturally die off, as it will no longer be able to sell their labour and hence acquire the means to survive. 

\emph{In this paper we will defend a somewhat opposite hypothesis -  the proliferation of intelligent machines will actually increase the number of labour hours, at least until the point where one can get machines that can replace men in their totality and reproduce as cheaply as humans.}

The rest of the paper is organised as follows: we gradually build a case for a historical trend in increasing working ours in section~\ref{hc}, while we envision a possible future in Section~\ref{tc}. We close with a short discussion in section~\ref{disc}.

\section{Machines and Labour}\label{hc}

\subsection{Job Automation} \label{sub:ja}
The introduction of ever-more automated machinery into production has been identified quite early on. Aristotle claimed that humanity will be forever in need of slaves, unless machines come to be that could \emph{anticipate} their master's desires~\cite{Aristoteles}. Chapter 5 of Marx's Capital is full of anecdotes and commentary on the use of automation in manufacturing and how it displaces and disciplines workers, using vocabulary that would be instantly recognisable to anybody doing research in the field today. More recently~\cite{bright1958automation}, we have seen scales of automation that take into account Aristotele's element of anticipation. Indeed, if you combine human-like intelligence and action versatility with reproduction costs lower than a human, there is a  case for an ``exterminism'' scenario - though there is still an argument to be made about machines that are fully obedient~\cite{caffentzis1999end,caffentzis1997machines}. Given that it is not clear when such human replacements will be available\footnote{The author speculates that machines with human-like qualities are at least 100 years away.}, what is a more pressing questions is to try and speculate what happens to humanity before this end-point is reached.

\begin{table}
\label{table:levels}
\begin{tabularx}{\textwidth}{ccXP{3cm}P{2cm}}
Group & Level & Level of mechanisation & Control Type & Type of machine response \\
 \toprule
 
\multirow{3}{*}{1} & 20 & Creates social needs and fulfils them & \multirow{3}{*}{\thead{Game-theoretic \\ Creative}} & \multirow{3}{*}{\thead{Social \\ actions}} \\
& 19 & Anticipates social needs and adapts  \\

& 18 & Anticipates human needs and adapts accordingly on social contexts   \\

  \midrule

\multirow{4}{*}{2} & 17 & Anticipates action required and adjusts to provide it & \multirow{4}{*}{\thead{Closed-loop \\ Online Planning }} & \multirow{4}{*}{\thead{Closed \\ large \\ actionset}} \\

& 16 & Corrects performance while operating    \\
& 15 & Corrects performance after operating  \\
  \midrule

\multirow{6}{*}{3} &  14 & Identifies and selects appropriate set of actions & \multirow{6}{*}{\thead{Partially \\ Closed-loop }} & \multirow{6}{*}{\thead{Fixed \\ actions}} \\  
& 13 & Segregates or rejects according to measurement \\ 
& 12 & Changes speed, position, direction according to measurement signal \\ 
\midrule
\multirow{4}{*}{4} & 11 & Records performance  & \multirow{4}{*}{\thead{Open-loop}} &  \multirow{4}{*}{\thead{Closed \\ actionset \\ Sensor \\ outputs}} \\  
& 10 & Signals preselected values of measurement (includes error detection) \\ 
& 9 & Measures characteristic of work \\
\midrule
\multirow{6}{*}{5} & 8 & Actuated by introduction of work piece or material & \multirow{6}{*}{\thead{Man \\ Open-Loop}} & \multirow{6}{*}{\thead{Fixed \\ actions} }\\  

& 7 & Power-tool system, remote controlled \\
& 6 & Power-tool, program control (sequence of fixed functions) \\
& 5 & Power-tool, fixed cycle (single function) \\ 
 \midrule
 
\multirow{4}{*}{6} & 4 & Power tool, hand control & \multirow{4}{*}{Man} & \multirow{4}{*}{Variable} \\
& 3 & Powered hand tool \\ 
& 2 & Hand tool \\ 
& 1 & Hand \\
 \bottomrule
\end{tabularx}
    
\caption{Levels of Automation - the argument raised here is that the higher the automation level, the more a machine looks like a plant or a beast of burden - reaching human like qualities only at the very top.}
\end{table}

In order to better define the quality of machines that are to be expected, we will use the Bright's scale~\cite{bright1958automation}, as quoted in \cite{braverman1998labor}. We present a modernised version of the scale with 3 more additions (group 1/ levels 18,19,20) in table \ref{table:levels}, to reflect machines with social mastery abilities. Current factory automation is usually at Group 4 - they perform some routine tasks without taking any kind of feedback into account - possibly with some newer factories at Group 3. A large amount of labour in modern societies happens at level 6 and 5, i.e. human controlled tools and machines that perform repetitive motions. 

Apart from the intelligence/automation level, one has to also take into account the reproductive qualities of machines. Designing machines that can self-reproduce is hard to be imagined when the overall technology level is at level 6. From level 2 onwards, it is possible to imagine machines that have elaborate control mechanisms that can create copies of themselves\footnote{Note that a marketeer that uses some kind of software to perform analysis and A/B tests uses a level 6 tool. }. This would be the equivalent of a factory machine that bears drills, ovens etc, effectively turning manufacturing into a form of farming. 
Machines of capabilities up to group 4 are indeed in direct competition with workers for wages. On an equal footing, all a worker can do is compete on \emph{time} and \emph{adaptability}; more office and managerial positions, jobs resulting from entertainment (e.g. mining gold in virtual environments for rich patrons, pornography, reality shows)~\cite{graeber2013phenomenon}. This can be combined with more traditional jobs in mining and textiles where automation is possible, but the labour costs so low that it is pointless to replace anyone. On par with the above, a trend can also be seen in the creation of insecure, low paid jobs with no career prospects, dubbed the secondary labour market~\cite{dickens1986labor}.

\subsection{Labour Hours}
Identifying historical trends in labour hours among different societal strata is hard, given the scarcity of data in the field~\footnote{See \url{http://groups.csail.mit.edu/mac/users/rauch/worktime/hours_workweek.html} for some discussion and some very relevant references.}. What one can deduce easily however is that there was a massive increase of labour time during the two industrial revolutions. The first industrial revolution extends from  1760 up to 1840, while the second one extends from 1840 up to 1914 and the beginning of the Great War. Both eras brought a significant change in working hours; pre-industrial revolution peasants increased their labour time \cite{blanchard1978labour}
from 180 days per year and 11 hours per day, to almost full year working and a 69-hour week and after the first revolution~\cite{woytinsky1935hours}. The almost genocidal effects of working hours were mitigated by a set of laws during the second industrial revolution, which initially limited hours of work to 12 per day for British factory workers and to 8 hours per day for US Federal employees. Starting from what can be considered the beginning of the second industrial revolution (1870), data shows a consistent drop of working hours~\cite{huberman2007times}. This trend indeed lasted until 1980, when we observe a rise in working hours again, especially in the new world~\cite{huberman2007times}. 

This tendency of decreasing work hours from 1870 onwards led important economic thinkers to believe that further reduction were ahead, with Keynes famously predicting a four-hour working day (the equivalent of a 20-hour working week). This prediction never materialised, although arguably the material conditions (e.g. worker productivity, technological innovations etc) have all shown significant gains. In the socialist world, the same set of optimism permeated the economic thinking, with soviet leadership~\cite{stalin1952economic} aiming for (but never delivering) 5 hour workdays. The crisis of 1970s, the advent of the internet, lifestyle choices, the collapse of the Soviet Union and the further collapse of the labour movement have proved almost impossible to predict~\cite{Friedman2015}.

\subsection{Money Power}

So what do the the upper echelons of society do with the labour-power and natural resources they consume? A part is arguably spent on maintaining idleness and in the consumption and creation of rare delicacies~\cite{veblen1899theory,kapur2016plutonomy}, but this seems a rather degenerate state. A much more important aspect is the reshaping of society, and thus reality, in their own desires, arguably playing the role of Level 1 machines. This transformation of labour-power to social power forms the core of our societies, and there is no evidence that any change is to expected in this front. Furthermore, the drive for more automation is going to be intensified, as it plays nicely with the power fantasies, ambitions and fetishes of certain elites, while it actually increases and transforms production - as has done for centuries.

\subsection{Escape to Nature}
Agricultural production is no longer something a large part of humanity can or aspires to do - and we have seen a decline in agricultural employment~\cite{simon1995state}, a trend more pronounced in advanced economies. In practical terms, this means an inability of large parts of the population to both deal and get involved in a direct exchange with nature. Further improvements in agricultural robots will hasten this trend.

\section{The Case for Increased Working Hours} \label{tc}
We have seen four different historical trends/facts : a) automation has been identified as a normal socio-economic process, b) working hours have been going up, with no labour movement to act as a counter-balance,  c) work is still needed and there is going to be a push for more automation and d) there is no longer the possibility of an escape towards subsistence farming. To put it in other words, the current situation is not unique, there is a trend, there are reasons to believe the it will continue and there is no obvious way to stop it. 

Given our arguments above, we think the logical outcome of improved automation is the increase in working hours, a continuation of an existing trend. The effect might be as dramatic as during the first industrial revolution, with the same adverse societal effects, while at the same time bringing global production to new heights.

\section{Discussion}
\label{disc}
There is a very clear case to be made that the most probable future for labour is more labour, in direct competitions with machines for survival. Adverse health conditions related to overwork will become common place (e.g. see~\citeA{Overwork}), while businesses will advertise for 70+ work-hour weeks as normal\footnote{This is true even in the AI community - see an online discussion here~\url{https://np.reddit.com/r/recruitinghell/comments/7003zb/strong_work_ethic/}}. It is highly unlikely that anything can revert this trend, but two possible scenarios still exist. The first scenario is war at a global scale, where productive forces are destroyed to such an extent that rebuilding and rethinking constitutes a necessity. Given the existence of nuclear weapons, something like this is highly unlikely. The second scenario is some abrupt change in social relationships, which cannot be predicted at the moment - movements that call for a reduction in working hours are fringe at best, while the dominance of capital over labour is almost absolute.

\vskip 0.2in
\bibliography{sample,time}
\bibliographystyle{theapa}

\end{document}